\documentclass{article}
\usepackage{spconf,amsmath,graphicx,hyperref}
\usepackage{amssymb}
\usepackage{pifont}
\usepackage{booktabs}
\usepackage{graphicx} 
\usepackage{multirow}
\usepackage{cite}
\usepackage{subcaption} 
\usepackage{tabularx}
\usepackage{array}
\usepackage{ragged2e}
\usepackage{makecell}
\usepackage{booktabs}
\usepackage{underscore}
\usepackage{etoolbox}
\usepackage{url}
\usepackage{threeparttable}
\usepackage{hyperref}
\usepackage{tablefootnote} 
\usepackage{xurl}

\graphicspath{{fig/}}

\title{SingMOS-Pro: An Comprehensive Benchmark for \\ Singing Quality Assessment}
\name{Yuxun Tang$^{1}$, Lan Liu$^{2}$, Wenhao Feng$^{1}$, Yiwen Zhao$^{3}$, Jionghao Han$^{3}$, Yifeng Yu$^{4}$, Jiatong Shi$^{3}$, Qin Jin$^{1\dagger}$\thanks{$\dagger$ Corresponding author.}}

\address{$^{1}$ Renmin University of China \quad
         $^{2}$ Sun Yat-sen University \\
         $^{3}$ Carnegie Mellon University \quad
         $^{4}$ Georgia Institute of Technology \\
         \small{\texttt{\{tangyuxun, qjin\}@ruc.edu.cn}\quad
        \texttt{jiatongs@cs.cmu.edu}}
        }

\begin{document}
\ninept
\maketitle
\begin{abstract}
Singing voice generation progresses rapidly, yet evaluating singing quality remains a critical challenge. Human subjective assessment, typically in the form of listening tests, is costly and time consuming, while existing objective metrics capture only limited perceptual aspects. 
In this work, we introduce \textbf{SingMOS-Pro}, a dataset for automatic singing quality assessment. Building on our preview version SingMOS, which provides only overall ratings, SingMOS-Pro  extends the annotations of the additional data to include \textit{lyrics}, \textit{melody}, and \textit{overall} quality, offering broader coverage and greater diversity. The dataset contains 7,981 singing clips generated by 41 models across 12 datasets, spanning from early systems to recent state-of-the-art approaches. Each clip is rated by at least five experienced annotators to ensure reliability and consistency.
Furthermore, we investigate strategies for effectively utilizing MOS data annotated under heterogeneous standards and benchmark several widely used evaluation methods from related tasks on SingMOS-Pro, establishing strong baselines and practical references for future research. The dataset is publicly available at \url{https://huggingface.co/datasets/TangRain/SingMOS-Pro}.

\end{abstract}

\section{Introduction}
\label{sec:intro}

Singing voice generation has attracted wide interest from both academia and industry. This task aims to produce high-quality vocal tracks from inputs such as musical scores or source vocals, with an emphasis on accuracy, personalization, and expressiveness. It encompasses sub-tasks including singing voice synthesis (SVS), singing voice conversion (SVC), and singing voice resynthesis (SVR). Recent advancements powered by models such as RNN~\cite{shi2021naivernn}, Transformers~\cite{lu2020xiaoicesing, wu2024toksing}, Generative Adversarial Networks (GANs)~\cite{kong2020hifigan, Yamamoto2020pwg, tang2024singomd}, variance autoencoder~(VAE)~\cite{zhang2022visinger, visinger2, yu2024visinger2+} and diffusion models~\cite{liu2022diffsinger, zhang2024stylesinger, zhang2024tcsinger} have achieved remarkable improvements in synthesis quality. Despite these successes, most research efforts have focused primarily on advancing generation techniques, while the equally crucial area of singing quality assessment~(SQA) remains underexplored. Yet, SQA plays a vital role in systematically identifying deficiencies in current models and providing clear directions for improvement.

In singing voice generation tasks, the conventional SQA method combines both human subjective assessment and objective assessments. Human subjective evaluation, typically in the form of Mean Opinion Score (MOS), is widely regarded as the "gold standard" for SQA~\cite{saeki22utmos}. However, conducting standard subjective tests is time-consuming and labor-intensive, and their results often lack comparability across different experiments. On the other hand, existing objective metrics, such as mel-cepstral distortion, have shown a weak correlation with perceived audio quality~\cite{saeki22utmos, reddy2021dnsmos}. This discrepancy makes the rapid and accurate evaluation of singing vocals challenging. Consequently, there is an urgent need for an efficient, reliable, and universal SQA method for singing voice generation.

In the fields of speech generation, quality assessment faces similar issues, but automatic prediction models have emerged as effective solutions. Notably, systems such as  UTMOS~\cite{saeki22utmos} and DNSMOS~\cite{reddy2021dnsmos} have been widely adopted for rapid evaluation. In contrast, research on SQA remains relatively limited, with the most critical bottleneck being the lack of suitable datasets. To address this gap, we present SingMOS-Pro, the first multilingual and multi-task-focused MOS dataset for SQA.
SingMOS-Pro consists of our preview version SingMOS~\cite{tang2024singmos} together with an extended collection of additional data. SingMOS focuses on samples generated from SVS and SVC tasks and provides overall MOS annotations for each clip, with five professional annotators per sample. The extended part incorporates samples from recent SVS, SVR and song generation systems, with each clip annotated by five annotators for three dimensions: lyrics score, melody score and overall MOS score. In total, SingMOS-Pro contains 7,981 clips, covering  3,425 SVS clips, 1,307 SVC clips, 2,671 SVR clips, and 578 ground-truth samples.
The preview version has also been adopted as the Singing Track of the ASRU 2024 VoiceMOS Challenge~\cite{huang2024voicemos}, promoting MOS prediction for SVS and SVC tasks and advancing research in SQA.
In this work, we further discuss training set utilization in SQA and benchmark widely used assessment methods on SingMOS-Pro, establishing strong baselines and practical references for future studies.

Our contributions are summarized as follows: we introduce SingMOS-Pro, the first multilingual and multi-task-focused MOS dataset for SQA, which contains 7,981 clips from SVS, SVC, SVR, and ground-truth systems annotated along three dimensions (overall, lyrics, melody); and we benchmark widely used assessment methods on SingMOS-Pro while analyzing training data utilization strategies, providing strong baselines and insights for future work on automatic singing quality assessment.

\section{SingMOS-Pro Dataset}
\label{sec:overview}

\subsection{Overview}

The SingMOS-Pro dataset is the first multilingual and broadly covered SQA corpus, consisting of the preview version SingMOS together with an extended collection of additional data. The preview version SingMOS was adopted as the dataset for Track 2 of the ASRU 2024 VoiceMOS Challenge, which aims to promote research progress in SQA, while the extended part serves as a supplement that broadens both the scale and diversity of the corpus.

\begin{figure}[!t]
    \begin{subfigure}[b]{0.23\textwidth}
        \centering
        \includegraphics[width=\textwidth]{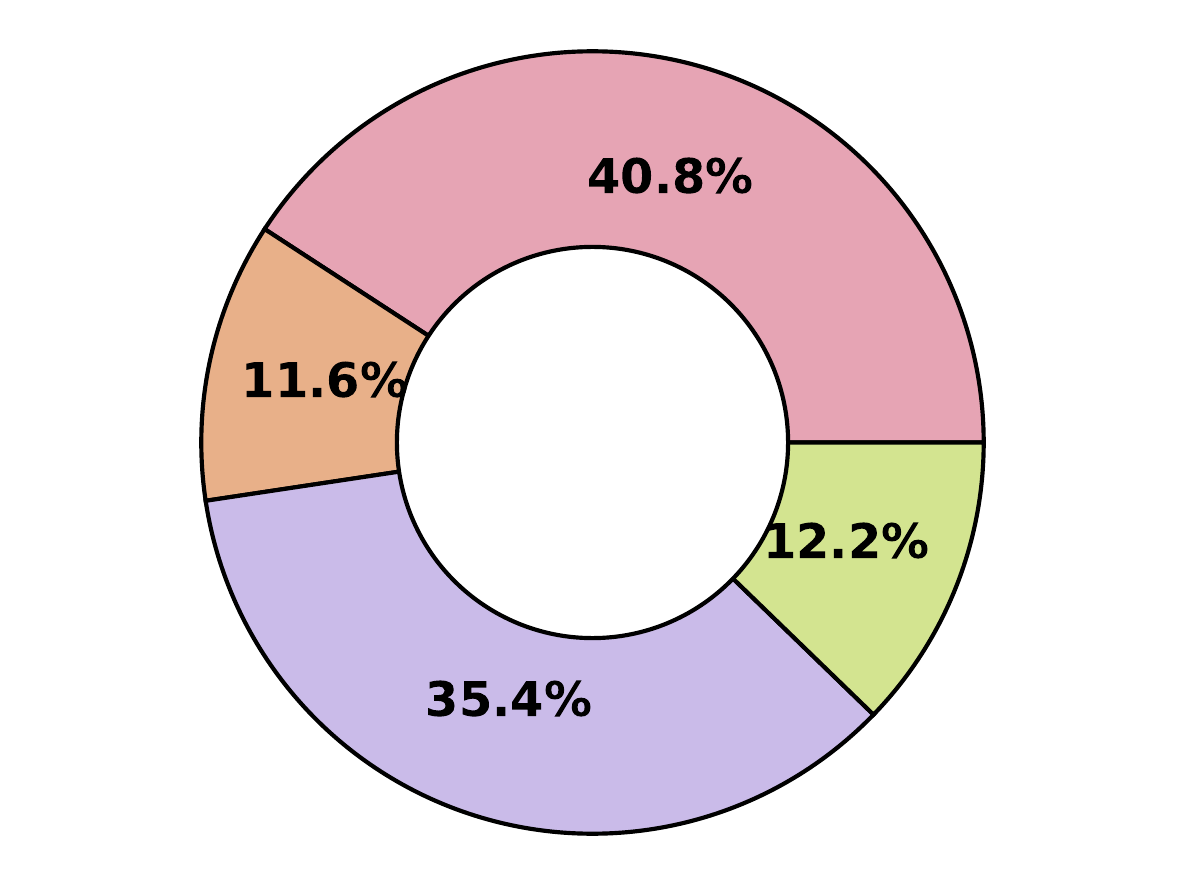}
        \caption{Systems Distribution}
        \label{fig:sys_distribution}
    \end{subfigure}
    \centering
    \begin{subfigure}[b]{0.23\textwidth}
        \centering
        \includegraphics[width=\textwidth]{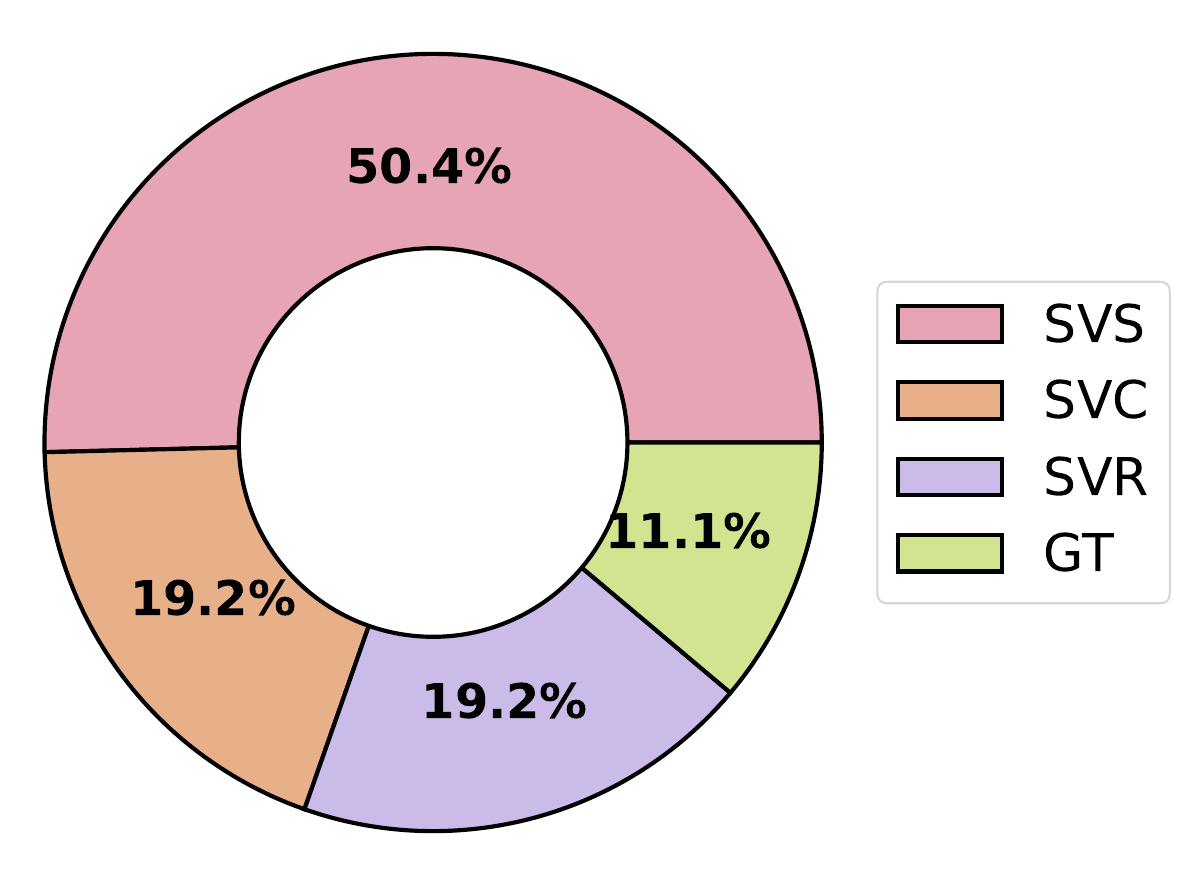}
        \caption{Utterances Distribution}
        \label{fig:utt_distribution}
    \end{subfigure}
    \hfill 
    \vspace{-5pt}
    \caption{Distribution of systems and utterances in the SingMOS-Pro. Subfigure (a) illustrates the distribution of systems across different tasks in the dataset, while subfigure (b) shows the distribution of utterances within each task.}
    \label{fig:utt_sys_distribution}
    \vspace{-10pt}
\end{figure}

For convenience, we define a system as the set of audio samples generated under a specific dataset–model–setting configuration. In total, SingMOS-Pro contains 11.15 hours of audio, consisting of 7,981 mono singing clips and 44,247 ratings collected from 78 experienced annotators. The distribution of clips is illustrated in Fig.~\ref{fig:utt_sys_distribution}: the dataset includes 3,425 clips from 60 SVS systems (plus 250 clips from song generation systems), 1,307 clips from 17 SVC systems, 2,671 clips from 52 SVR systems, and 578 ground-truth recordings from 12 systems. Among them, 6,937 clips are resampled to 16kHz, while 631 and 413 remain at 24kHz and 44.1kHz, respectively. Across all samples, the average clip duration is 5.03 seconds.
Each singing clip is evaluated by at least five experienced annotators for the overall MOS score. In addition, 4,155 clips are further annotated with lyrics scores and melody scores, which capture the clarity of pronunciation and the naturalness of melody, respectively.

More details regarding system and samples can be found in Section~\ref{sec:collection}, and the information about the annotation process is provided in Section~\ref{sec:annotation}.

\subsection{Statistics}
\label{sec: statistic}
\begin{figure}[!t]
    \centering
    \includegraphics[width=0.4\textwidth]{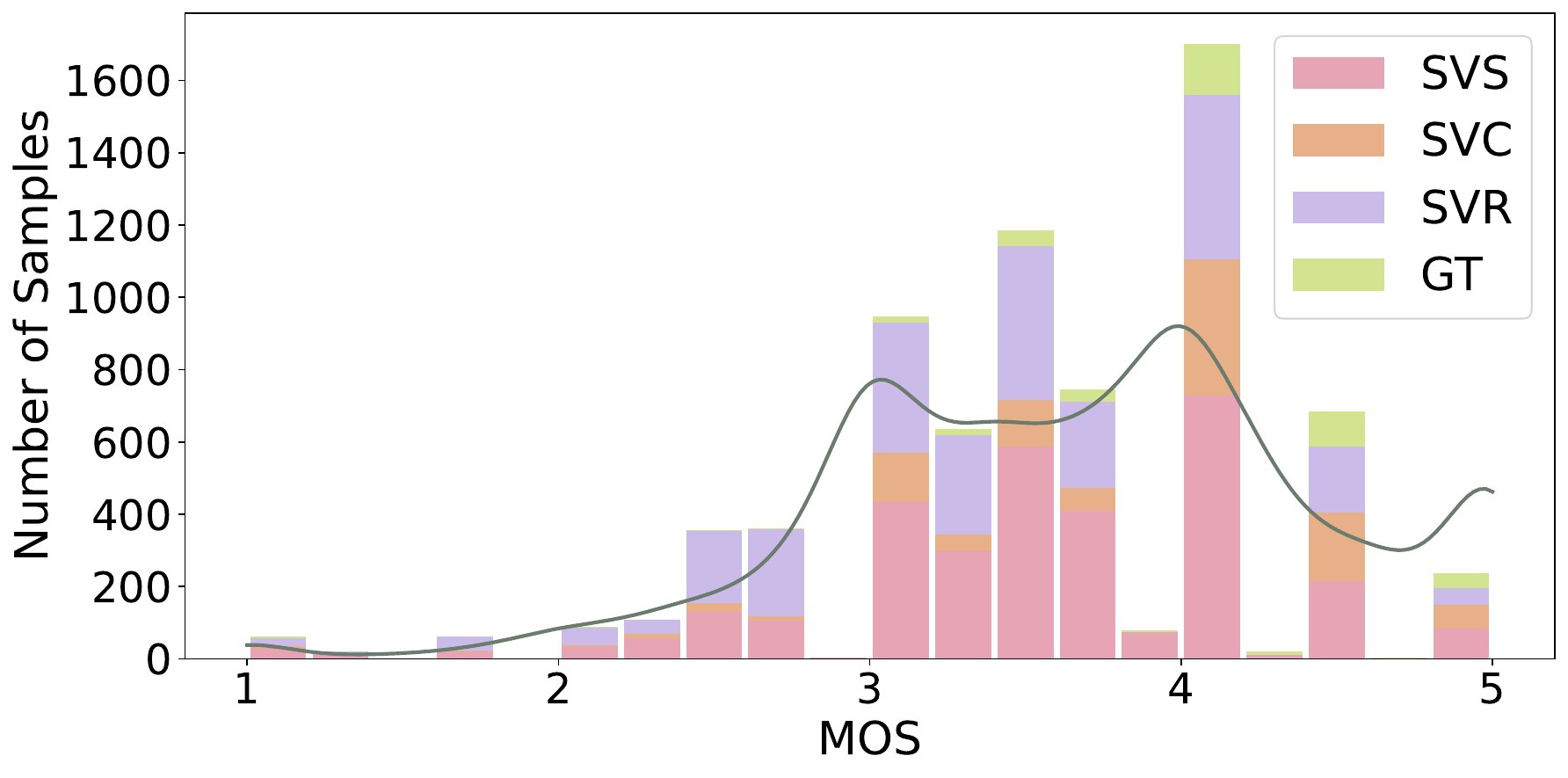}
    \vspace{-10pt}
    \caption{Distribution of Utterances Across MOS Intervals.}
    \label{fig:utt_mos_num}
    \vspace{-10pt}
\end{figure}

Fig.~\ref{fig:utt_mos_num} shows the distribution of utterances in SingMOS-Pro with respect to overall MOS scores. Most generated singing vocals are concentrated around scores between 3 and 4, with fewer samples between 2 and 3. In contrast, ground-truth recordings are mainly clustered between 4 and 5, while only a few noisy samples appear near 1. Overall, the distribution approximates a bimodal shape.

\begin{figure*}[!t]
    \centering
    \includegraphics[width=0.9\textwidth]{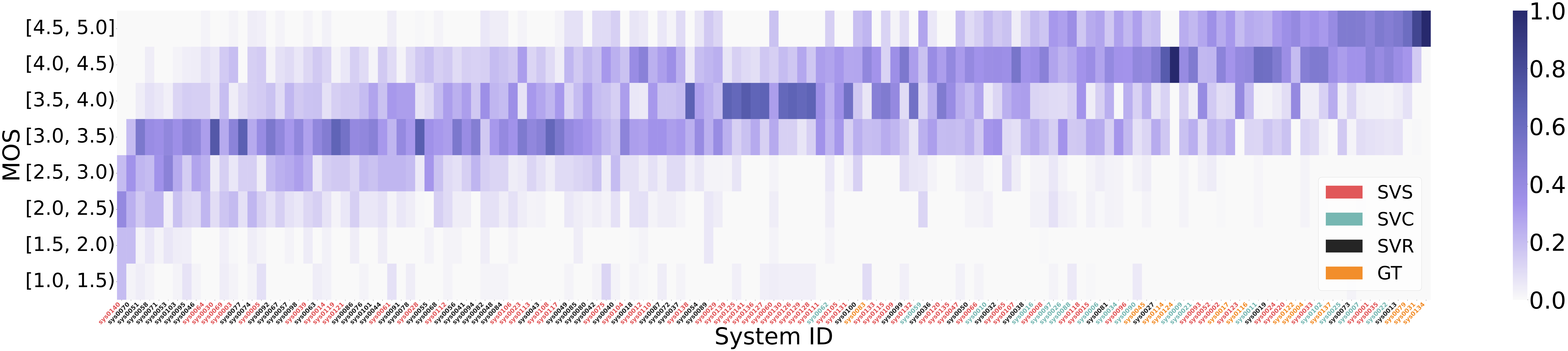}
    \vspace{-10pt}
    \caption{Distribution of Systems Across MOS Intervals. Detailed of \textit{System ID} is available at metadata in SingMOS-Pro.}
    \label{fig:sysmos}
    \vspace{-10pt}
\end{figure*}

Fig.~\ref{fig:sysmos} presents the MOS distributions across systems. Among the better-performing systems, SVR generally outperforms SVC, which in turn outperforms SVS, consistent with the expected hierarchy: SVR represents the upper bound of generation models, and SVC is typically easier to model than SVS. However, since SVR relies on speech-pretrained codec models, its performance degrades on singing datasets, leading to lower MOS scores. 

For the same configuration (VISinger2~\cite{visinger2} on Opencpop~\cite{wang2022opencpop} and DiffNNSVS~\cite{yamamoto2023nnsvs} on Namine\textsuperscript{1}) evaluated at different sampling rates, the performance ranks as 24kHz (3.88/4.03) $>$ 44.1kHz (3.81/3.98) $>$ 16kHz (3.65/3.80). Results at 24kHz and 44.1kHz are close, while both clearly outperform 16kHz. We assume that listeners can easily perceive the quality gap between 16kHz and 24kHz, whereas the slight drop at 44.1kHz may stem from limited modeling capacity.

\section{Dataset Construction}

\subsection{Singing Clips Collection}
\label{sec:collection}

To ensure diversity in the collected samples, we consider three aspects: dataset, model, and model setting.


\noindent \textbf{Dataset}: We collect 167 songs from the test sets of twelve publicly available singing datasets, covering over 50 singers. It includes five Mandarin datasets (Opencpop~\cite{wang2022opencpop}, M4Singer~\cite{zhang2022m4singer}, ACE-Opencpop~\cite{shi2024aceopencpop}, Kising~\cite{shi2024aceopencpop} and Chinese GTSinger~\cite{zhang2025gtsinger}) and seven Japanese datasets (JVSMusic~\cite{tamaru2020jvsmusic}, Kiritan Singing~\cite{kiritan}, 
Namine Ritsu\footnote{\scriptsize\url{https://drive.google.com/drive/folders/1XA2cm3UyRpAk_BJb1LTytOWrhjsZKbSN}},
Ofuton P\footnote{\scriptsize\url{https://sites.google.com/view/oftn-utagoedb/\%E3\%83\%9B\%E3\%83\%BC\%E3\%83\%A0}},
Ameboshi Cipher\footnote{\scriptsize\url{https://parapluie2c56m.wixsite.com/mysite}},
Natsume Singing\footnote{\scriptsize\url{https://github.com/AmanoKei/Natsume_Singing}} and
Oniku Kurumi\footnote{\scriptsize\url{https://onikuru.info/db-download/}}
). 
In addition, we construct a synthetic dataset by extracting melodies into musical scores with ChatMusician~\cite{yuan2024chatmusician} and generating lyrics with DeepSeek-V2~\cite{deepseekv2}.

\begin{table}[!t]
    \centering
    \footnotesize
    \caption{Models in SingMOS-Pro. \textit{Italicized models} indicate those evaluated under multiple settings. “X/Y” denotes the number of models X and the number of systems Y.}
    \vspace{-5pt}
    \setlength{\tabcolsep}{3pt}

    \begin{threeparttable}
        \begin{tabularx}{\linewidth}{@{} p{4.5em} X @{}}
            \toprule
            \textbf{Cate} & \textbf{Models} \\
            \midrule
            GT(1/12) & Ground-Truth \\
            \midrule
            SVC(2/17) & \textit{Sovits} \cite{yamamoto2023nusvc}, \textit{Nusvcc} \cite{yamamoto2023nusvc} \\
            \midrule
            SVR(9/52) & \textbf{Codec:} \textit{SoundStream} \cite{zeghidour2021soundstream}, \textit{Encodec} \cite{defossez2022high}, \textit{DAC} \cite{kumar2023DAC}; \\[2pt]
            & \textbf{Vocoder:} \textit{HiFiGAN} \cite{kong2020hifigan}, MelGAN \cite{kumar2019melgan}, Parallel WaveGAN \cite{Yamamoto2020pwg}, DiffWave \cite{kong2021diffwave}, WaveGrad \cite{chen2020wavegrad}, WaveNet \cite{vandenoord16wavenet} \\
            \midrule
            SVS \newline (26/60) & \textbf{Pretrained Model:} VISinger2 \cite{visinger2}, VISinger \cite{zhang2022visinger}, \textit{XiaoiceSing} \cite{lu2020xiaoicesing}, \textit{VISinger2+} \cite{yu2024visinger2+}, DiffNNSVS \cite{yamamoto2023nnsvs}, RNN \cite{shi2021naivernn}, Diffsinger \cite{liu2022diffsinger}, \textit{svsAug} \cite{zhao2025svsaug}, StyleSinger \cite{zhang2024stylesinger}, TCSinger \cite{zhang2024tcsinger}, TechSinger \cite{guo2025techsinger}, ACE Studio \cite{shi2024aceopencpop}, \textit{SingOMD} \cite{tang2024singomd}, \textit{TokSing} \cite{wu2024toksing}, Sinsy \cite{yamamoto2023nnsvs}, ARNNSVS \cite{yamamoto2023nnsvs}; \\[2pt]
            & \textbf{Official Demo:} ExpressiveSinger\tnote{a}, SPSinger\tnote{b}, PeriodSinger\tnote{c}, EveryoneCanSing\tnote{d}, TCSinger2\tnote{e}, XSinger\tnote{f}; \\[2pt]
            & \textbf{Song Generation:} ACEStep \cite{gong2025acestep}, Diffrhythm \cite{ning2025diffrhythm}, Hailuo\tnote{g}, Suno\tnote{h}, Yue \cite{yuan2025yue} \\
            \bottomrule
        \end{tabularx}

        \begin{tablenotes}
            \scriptsize 
            \item[a] \url{https://expressivesinger.github.io/ExpressiveSinger/}
            \item[b] \url{https://danny-nus.github.io/SPSinger/}
            \item[c] \url{https://rlataewoo.github.io/periodsinger/}
            \item[d] \url{https://everyone-can-sing.github.io/}
            \item[e] \url{https://aaronz345.github.io/TCSinger2Demo/}
            \item[f] \url{https://jisang93.github.io/x-singer/}
            \item[g] \url{https://hailuoai.com/music}
            \item[h] \url{https://suno.com/}
        \end{tablenotes}

    \end{threeparttable}

    \vspace{-10pt}
    \label{tab:system_categories}
\end{table}

\noindent \textbf{Model and Model Setting}: We adopt a wide range of SVS, SVC, and SVR models, covering diverse architectures and configurations. All open-source models are either official pre-trained versions or trained using toolkits such as ESPnet~\cite{wu2024muskits, shi22muskits, espnet_codec}, NNSVS~\cite{yamamoto2023nnsvs}, and vocoder-benchmark~\cite{albadawy2022vocbench}. Table~\ref{tab:system_categories} summarizes all models, with those in bold indicating those evaluated under multiple settings. Additional metadata is provided in the released dataset.

In practice, part of the generated samples is directly obtained from the CtrSVDD Challenge 2024~\cite{zang2024ctrsvdd}, which follows the same procedure.
For models not publicly available (Demo in Table.~\ref{tab:system_categories}), we supplement SingMOS-Pro with samples collected from their official demo pages. For song generation models, we synthesize samples with their demo models in the pop style using Chinese lyrics from Opencpop and GTSinger, and apply MelBand Roformer~\cite{wang2023melbandroformer} to separate the vocals from the accompaniment.

\begin{figure}[!t]
    \begin{subfigure}[t]{0.23\textwidth}
        \centering
        \includegraphics[width=\textwidth]{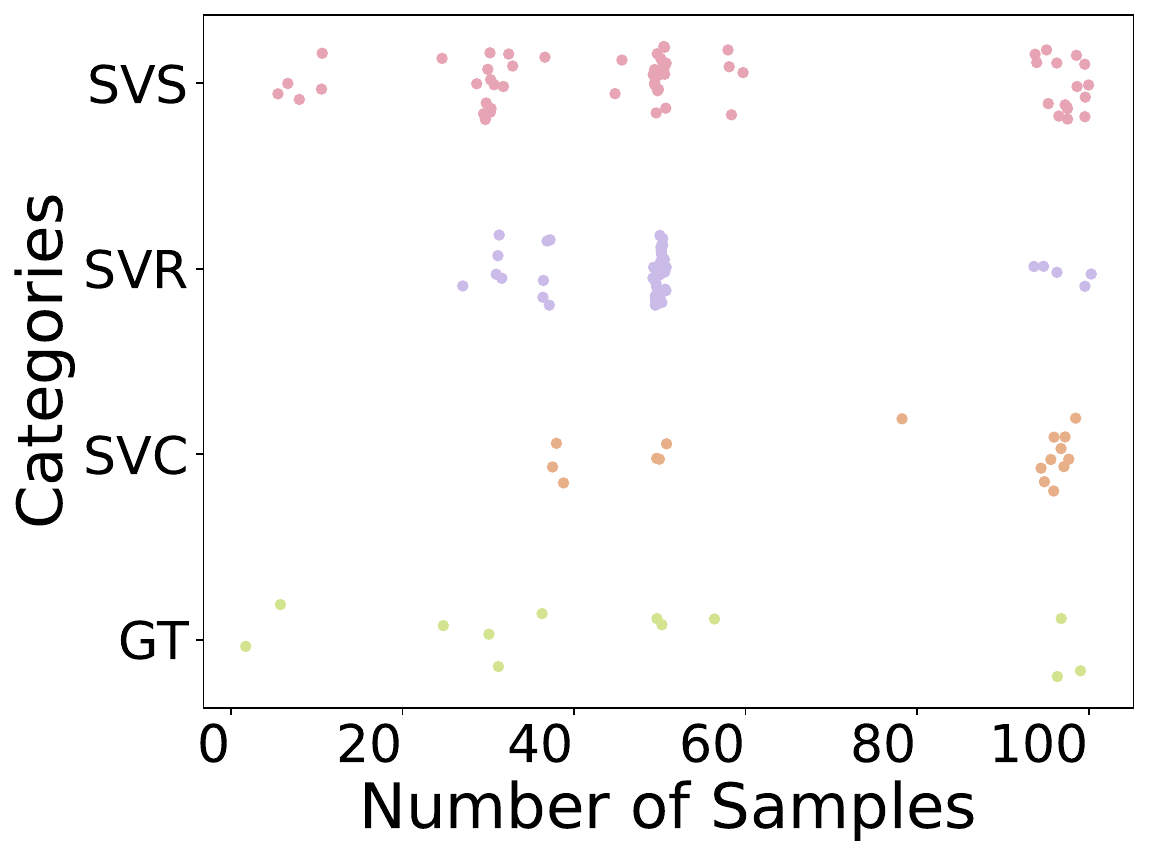}
        \caption{Utterance Number of Systems}
        \label{fig:sys_utt_num}
    \end{subfigure}
    \centering
    \begin{subfigure}[t]{0.23\textwidth}
        \centering
        \includegraphics[width=\textwidth]{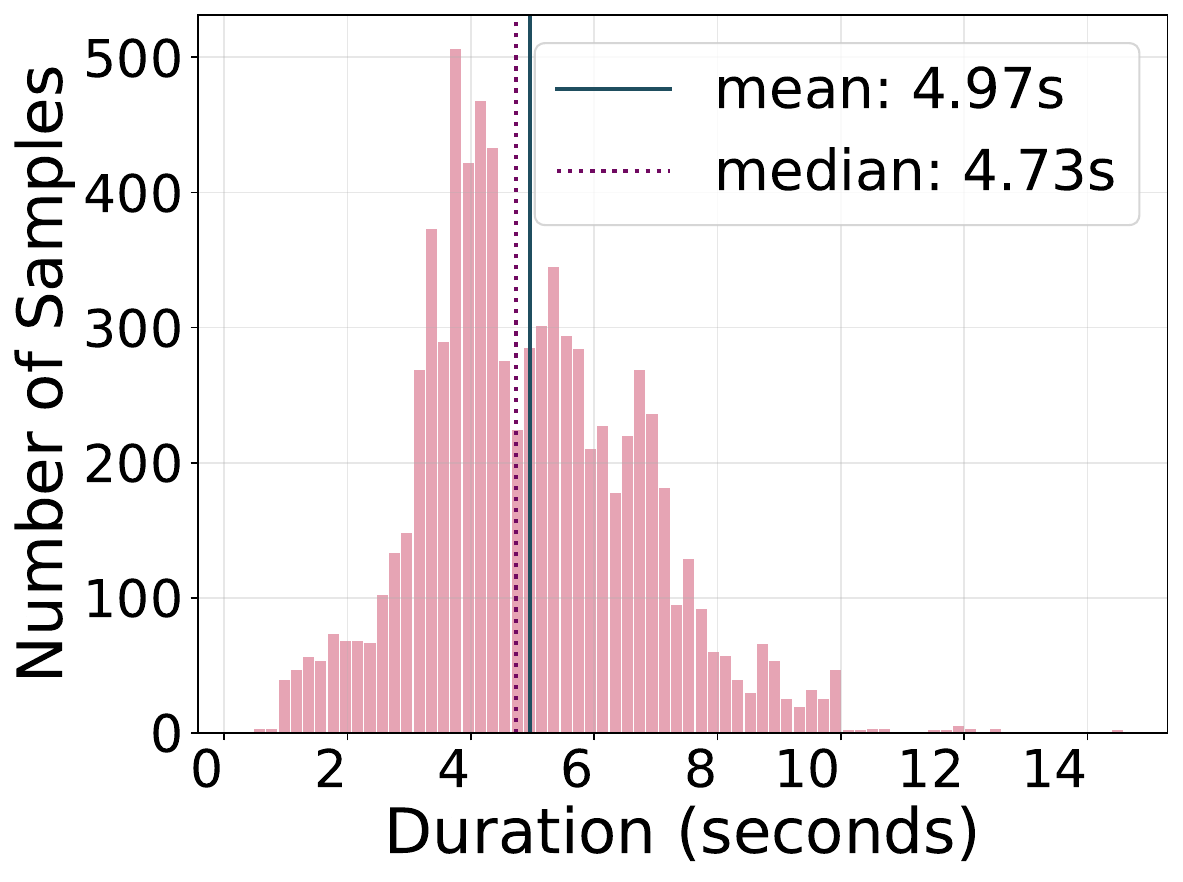}
        \caption{Utterance Duration Distribution}
        \label{fig:utt_dur_distribution}
    \end{subfigure}
    \hfill 
    \vspace{-6pt}
    \caption{Utterances Composition Statistics (few extreme outliers are discarded in visualization). Subfigure (a) presents the number of utterances in each system, and subfigure (b) depicts the distribution of utterance durations.}
    \label{fig:utt_per_system}
    \vspace{-10pt}
\end{figure}

In total, we collect 141 systems, producing 7,981 clips comprising both synthetic and real singing vocals. Fig.~\ref{fig:utt_per_system} illustrates the distribution of clips per system and the duration of utterances.

\subsection{Annotation Protocols}
\label{sec:annotation}
In SingMOS-Pro, we recruite 78 annotators to conduct the evaluation. In total, 7,981 audio clips are annotated, yielding 44,247 overall-performance ratings. Among these, 4,155 clips are additionally annotated for lyric and melody, resulting in 23,475 lyric ratings and 23,475 melody ratings.

\noindent \textbf{Evaluation Dimensions:} Lyrics MOS specifically measures pronunciation clarity and intelligibility, while Melody MOS evaluates melodic naturalness and pitch accuracy. Finally, Overall MOS provides a holistic assessment, capturing the general quality by considering both lyrical and melodic dimensions.

\noindent \textbf{Evaluation Design:} All MOS tests are conducted online. ensure data quality, all annotators are required to complete a pre-annotation training lecture. Evaluations are performed in quiet environments, and the audio samples are annotated in five discrete batches over time. Each audio clip receives at least five independent annotations. The first and forth batch collect only overall MOS, while the second, third and fifth batches include three dimensions: overall, lyrics, and melody scores. The fourth and fifth batches are annotated for other projects~\cite{zhao2025svsaug, yu2024visinger2+}. All scores are collected on a 5-point Likert scale. Notably, the preview version SingMOS~\cite{tang2024singmos} consists entirely of samples from the first batch.

\noindent \textbf{Quality Control:} Although human judgment is regarded as the gold standard for audio evaluation, issues such as distraction and fatigue may introduce mistakes. To mitigate this, each batch contained trap clips (noise or silence) and carefully selected golden clips. If an annotator assigned high scores to trap clips or low scores to golden clips, the entire batch will be re-annotated.

\subsection{Dataset Splits}


\label{sec:split}
To facilitate the effective use of our dataset, we predefine a train/test split. For the first, second, and third annotation batches, systems  more than 50 clips are split into training and test sets with a 7:3 ratio. Systems containing 10–50 clips are entirely assigned to the test set, while systems with fewer than 10 clips are included only in the training set. For the fourth and fifth annotation batches, all results are merged into the training set. Since annotation standards vary across batches, we maintain separate test sets for different batches. The impact of these decisions on the training set will be further discussed in Section~\ref{sec: train_set}. After the split, the training set contains 4,453 clips, while \textit{test1}, \textit{test2}, and \textit{test3} consist of 1 070, 1 444, and 339 clips, respectively. 

\begin{table*}[!!t]
\centering
\footnotesize
\caption{Performance of Different Settings Using Training Dataset. 
Results are reported in the format ``R1/R2$|$R3$|$Avg'', aligned to the outcomes on \textit{test1}, \textit{test2}, \textit{test3}, and their weighted average, where \textbf{D.id} indicates domain id and \textbf{MDF} denotes multi-dataset finetuning.}
\vspace{-10pt}
\begin{tabular}{c c  ccc  ccc}
\toprule 
\multirow{2}{*}{\textbf{D.id}} & 
\multirow{2}{*}{\textbf{MDF}} &
\multicolumn{3}{c}{\textbf{Utterance-level}} & 
\multicolumn{3}{c}{\textbf{System-level}} \\
\cmidrule(lr){3-5} \cmidrule(lr){6-8}
& & \textbf{RMSE}$\downarrow$ & \textbf{LCC}$\uparrow$ & \textbf{SRCC}$\uparrow$
& \textbf{RMSE}$\downarrow$ & \textbf{LCC}$\uparrow$ & \textbf{SRCC}$\uparrow$ \\
\midrule
\ding{55}    & \ding{55}       & 0.54$|$0.47$|$0.42$|$0.49 
                                & \textbf{0.49}$|$\textbf{0.72}$|$0.32$|$0.58 
                                & 0.47$|$0.57$|$0.30$|$0.50  
                                & 0.06$|$\textbf{0.07}$|$0.11$|$0.07  
                                & 0.86$|$\textbf{0.93}$|$0.61$|$0.86  
                                & 0.81$|$0.78$|$0.62$|$\textbf{0.77}  \\

$\checkmark$ & \ding{55}       & \textbf{0.48}$|$0.58$|$\textbf{0.30}$|$0.51 
                                & 0.54$|$0.67$|$\textbf{0.35}$|$0.58 
                                & \textbf{0.51}$|$0.52$|$\textbf{0.34}$|$0.50 
                                & 0.05$|$0.11$|$\textbf{0.05}$|$0.08 
                                & \textbf{0.90}$|$0.90$|$\textbf{0.68}$|$\textbf{0.87} 
                                & 0.82$|$0.71$|$\textbf{0.67}$|$0.74   \\

\ding{55}    & $\checkmark$    &  0.54$|$0.47$|$0.40$|$0.49 
                                & 0.51$|$\textbf{0.72}$|$0.28$|$0.58 
                                & 0.49$|$\textbf{0.58}$|$0.27$|$0.51 
                                & 0.05$|$\textbf{0.07}$|$0.09$|$0.07 
                                & 0.87$|$\textbf{0.93}$|$0.57$|$0.86 
                                & 0.78$|$\textbf{0.79}$|$0.57$|$0.76  \\

$\checkmark$ & $\checkmark$    & 0.49$|$\textbf{0.45}$|$0.31$|$\textbf{0.45} 
                                & 0.54$|$\textbf{0.72}$|$0.32$|$\textbf{0.60} 
                                & \textbf{0.51}$|$\textbf{0.58}$|$0.30$|$\textbf{0.52} 
                                & \textbf{0.04}$|$0.08$|$0.06$|$\textbf{0.06}
                                & \textbf{0.90}$|$0.92$|$0.62$|$\textbf{0.87 }
                                & \textbf{0.83}$|$0.76$|$0.55$|$0.75  \\
\bottomrule
\end{tabular}

\label{tab:train_set}
\end{table*}

\begin{table*}[!!t]
\centering
\footnotesize
\caption{Model Comparison on SingMOS-Pro at Utterance/System Levels. ``PM'' denotes MIDI pitch, ``PH'' denotes pitch histograms.}
\vspace{-10pt}
\resizebox{\linewidth}{!}{
\begin{tabular}{l c cccc cccc}
\toprule
\multirow{2}{*}{\textbf{Model}} & \multirow{2}{*}{\textbf{FineTune}} 
& \multicolumn{3}{c}{\textbf{Utterance-level}} 
& \multicolumn{3}{c}{\textbf{System-level}} \\
\cmidrule(lr){3-5} \cmidrule(lr){6-8}
 &  & \textbf{RMSE} & \textbf{LCC} & \textbf{SRCC} 
    & \textbf{RMSE} & \textbf{LCC} & \textbf{SRCC} \\
\midrule
DNSMOS                  & \ding{55} & 1.10$|$0.90$|$0.71$|$0.96 & 0.24$|$0.57$|$0.23$|$0.39 & 0.23$|$0.45$|$0.20$|$0.33 & 0.85$|$0.64$|$0.55$|$0.69 & 0.41$|$0.74$|$0.48$|$0.60 & 0.28$|$0.45$|$0.51$|$0.41 \\
UTMOS                   & \ding{55} & 1.96$|$1.94$|$1.75$|$1.93 & 0.35$|$0.19$|$0.16$|$0.26 & 0.35$|$0.43$|$0.12$|$0.36 & 1.78$|$1.78$|$1.69$|$1.77 & 0.66$|$0.18$|$0.39$|$0.36 & 0.49$|$0.60$|$0.45$|$0.54 \\
\textbf{}SingMOS                 & \ding{55} & 0.55$|$0.86$|$0.75$|$0.71 & 0.72$|$0.55$|$0.14$|$0.57 & \textbf{0.70}$|$0.48$|$0.09$|$\textbf{0.53} & 0.19$|$0.58$|$0.54$|$0.45 & \textbf{0.96}$|$0.79$|$0.32$|$0.78 & 0.95$|$0.63$|$0.35$|$0.69 \\
SHEET-ssqa              & \ding{55} & 0.63$|$0.74$|$0.79$|$0.70 & 0.59$|$0.65$|$0.11$|$0.56 & 0.59$|$0.52$|$0.07$|$0.50 & 0.20$|$0.38$|$0.50$|$0.34 & 0.89$|$0.85$|$0.34$|$0.79 & 0.81$|$0.68$|$0.43$|$0.69 \\
\midrule
SSL              & $\checkmark$ & 0.54$|$\textbf{0.47}$|$0.42$|$\textbf{0.49} 
                                & 0.49$|$\textbf{0.72}$|$0.32$|$\textbf{0.58} 
                                & 0.47$|$\textbf{0.57}$|$0.30$|$0.50  
                                & 0.06$|$0.07$|$0.11$|$\textbf{0.07}  
                                & 0.86$|$0.93$|$0.61$|$0.86  
                                & 0.81$|$0.78$|$0.62$|$0.77  \\
SSL+PM           & $\checkmark$ & 0.53$|$0.48$|$\textbf{0.41}$|$\textbf{0.49} 
                                & 0.51$|$0.71$|$\textbf{0.35}$|$\textbf{0.58} 
                                & 0.48$|$\textbf{0.57}$|$\textbf{0.33}$|$0.50 
                                & 0.05$|$0.08$|$\textbf{0.09}$|$\textbf{0.07} 
                                & 0.87$|$0.92$|$\textbf{0.63}$|$0.86 
                                & 0.79$|$0.76$|$\textbf{0.69}$|$0.76 \\
SSL+PH           & $\checkmark$ &  \textbf{0.51}$|$0.48$|$0.43$|$\textbf{0.49} 
                                & 0.52$|$0.71$|$0.31$|$\textbf{0.58} 
                                & 0.50$|$0.56$|$0.31$|$0.51 
                                & \textbf{0.04}$|$0.07$|$0.13$|$\textbf{0.07 }
                                & 0.90$|$\textbf{0.94}$|$0.58$|$\textbf{0.88} 
                                & 0.83$|$\textbf{0.82}$|$0.61$|$\textbf{0.79}  \\
\bottomrule
\end{tabular}}
\vspace{-10pt}
\label{tab:model_performance}
\end{table*}

\section{SQA Benchmark}
\label{sec:benchmark}

This section addresses the challenge of leveraging datasets from different annotation batches for SQA and presents a unified evaluation of representative methods on the SingMOS-Pro. 

\subsection{Experiments Setting}

All experiments are conducted on SingMOS-Pro with only 16kHz audio, since our self-supervised learning~(SSL) backbone supports only 16kHz inputs. All 24kHz and 44.1kHz samples are filtered out, resulting in 4,007 clips and 87 systems for training and 2,091/1,540/376 clips and 31/41/15 systems for \textit{test1}/\textit{test2}/\textit{test3}.
The SSL backbone model is wav2vec2-large~\cite{baevski2020wav2vec} and is optimized with L1 loss with margin, using stochastic gradient descent with a learning rate of 0.001 and a momentum of 0.9. Training is performed for 200 epochs with a batch size of 15. For evaluation, we report Root Mean Squared Error (RMSE), Linear Correlation Coefficient (LCC), and Spearman’s Rank Correlation Coefficient (SRCC). Among these metrics, SRCC is considered the most important, since it reflects the ranking consistency of perceptual quality.
To facilitate a comprehensive evaluation, we compute weighted averages of the results from \textit{test1}, \textit{test2}, and \textit{test3} at both the utterance and system levels, where the weights are determined by the corresponding number of utterances and systems in each test set.

\subsection{Exploration on Train Set Utilization}
\label{sec: train_set}

In audio quality assessment, whether for speech or singing, it is challenging to collect a sufficiently large and diverse dataset under a unified annotation standard. This makes it crucial to effectively leverage data annotated with different criteria. Following the approaches in AlignNet~\cite{Pieper2024alignnet} and SHEET~\cite{huang2024sheet}, we investigate two strategies on a self-supervised learning backbone. The first strategy is multi-dataset finetuning (MDF), in which we first train on the training set of the first annotation batch for 10 epochs and then continue finetuning on the entire training set. The second strategy is the use of the domain id, which corresponds to the specific annotation batch index.

As shown in Table~\ref{tab:train_set}, the SSL model without any auxiliary design and the SSL model equipped with both MDF and domain id achieve the best weighted SRCC at the system level and utterance level, respectively. Adding domain id leads to clear improvements on \textit{test3}, suggesting that domain identifiers are particularly beneficial for test sets with fewer samples, as they help the model learn to mitigate domain discrepancies. Also, introducing MDF consistently improves performance on most metrics of \textit{test2}, indicating that multi-dataset finetuning alleviates the confusion caused by different annotation standards across batches. Considering the weighted results across all test sets, the joint use of domain ids and MDF yields the most effective overall performance, while directly using the plain SSL model also proves to be a reasonable and convenient choice.

\subsection{Exploration with Singing Quality Assessment}
In this experiment, we conduct a horizontal comparison of models under different settings on SingMOS-Pro to examine the impact of various factors on SQA. For the sake of experimental simplicity and to eliminate the interference of confounding factors, we adopt a plain SSL model as the backbone. We first compare several pretrained baselines, including the speech MOS models UTMOS~\cite{saeki22utmos} and DNSMOS~\cite{reddy2021dnsmos}, SingMOS trained on the preview version of SingMOS, and SHEET-ssqa~\cite{huang2024sheet} trained on both speech MOS data and the preview version of SingMOS.
As shown in the first four rows of Table~\ref{tab:model_performance}, speech MOS models perform poorly on the singing task due to the substantial domain gap. The SingMOS model trained on the preview version achieves the best SRCC on \textit{test1}, but its performance drops significantly on the out-of-domain \textit{test2} and \textit{test3}, indicating clear overfitting and the need for broader data coverage. In contrast, SHEET-ssqa, which integrates additional speech MOS data, alleviates the overfitting issue on out-of-domain sets to some extent, suggesting that combining speech and singing MOS data is a promising direction. 

Since SQA requires greater emphasis on melodic naturalness compared to speech assessment, we further explore several strategies for using pitch information. These strategies include the pitch histogram proposed in~\cite{shi2024pitchmos}, as well as MIDI pitch and MIDI pitch variance, which are extracted from the raw waveform and quantized by $f_0$-to-MIDI conversion. As shown in the last three rows of Table~\ref{tab:model_performance}, the pitch histogram yields slightly better performance than MIDI pitch, but the overall improvement over the SSL baseline remains marginal, underscoring the need for further exploration of how melodic cues can be more effectively integrated. Beyond these approaches, exploring how to incorporate melody scores and lyric scores into SQA constitutes a valuable avenue for future work.

\section{Conclusion}
We introduced SingMOS-Pro, the first multilingual, multi-task, and fine-grained MOS dataset for automatic SQA. By providing reliable annotations across lyrics, melody, and overall dimensions, and by benchmarking widely used evaluation methods, SingMOS-Pro establishes strong baselines and offers practical references for future research. We believe SingMOS-Pro will facilitate the development of more effective and robust SQA models. Beyond these approaches, exploring how to incorporate melody scores and lyric scores into SQA constitutes a valuable avenue for future work.


\bibliographystyle{IEEEbib}
\bibliography{strings}

\end{document}